\documentclass[%
 reprint,
 amsmath,amssymb,
 aps,
floatfix,
]{revtex4-2}

\usepackage{amsmath}

\usepackage{xcolor}
\usepackage{graphicx}
\usepackage{dcolumn}
\usepackage{bm}
\usepackage{amsmath}
\usepackage{enumitem}
\usepackage{xcolor}
\usepackage{graphicx}
\usepackage{dcolumn}
\usepackage{bm}
\usepackage{float}

\begin{document}

\preprint{APS/123-QED}


\title{
Network Representation of Higher-Order Interactions Based on Information Dynamics
}


\author{Gorana Mijatovic}
 \altaffiliation{Faculty of Technical Sciences, University of Novi Sad, Serbia}
 \email{gorana86@uns.ac.rs}

 \author{Yuri Antonacci}
 \altaffiliation{Department of Engineering, University of Palermo, Italy}
 \email{yuri.antonacci@unipa.it }


 \author{Michal Javorka}
 \altaffiliation{Department of Physiology and Biomedical Center Martin, Jessenius Faculty of Medicine, Comenius University, Slovakia}
 \email{michal.javorka@uniba.sk}

 \author{Daniele Marinazzo}
 \altaffiliation{Department of Data Analysis, University of Ghent, Belgium}
 \email{Daniele.Marinazzo@ugent.be }

\author{Sebastiano Stramaglia}
 \altaffiliation{Department of Physics, University of Bari Aldo Moro, and INFN Sezione di Bari, Italy}
 \email{sebastiano.stramaglia@uniba.it}
 
\author{Luca Faes}
\altaffiliation{Department of Engineering, University of Palermo, Italy}
\email{luca.faes@unipa.it}


\begin{abstract}

Many complex systems in science and engineering are modeled as networks whose nodes and links depict the temporal evolution of each system unit and the dynamic interaction between pairs of units, which are assessed respectively using measures of auto- and cross-correlation or variants thereof.
However, a growing body of work is documenting that this standard network representation can neglect potentially crucial information shared by three or more dynamic processes in the form of higher-order interactions (HOIs).
While several measures, mostly derived from information theory, are available to assess HOIs in network systems mapped by multivariate time series, none of them is able to provide a compact and detailed representation of higher-order interdependencies.
In this work, we fill this gap by introducing a framework for the assessment of HOIs in dynamic network systems at different levels of resolution. The framework is grounded on the dynamic implementation of the O-information, a new measure assessing HOIs in dynamic networks, which is here used together with its local counterpart and its gradient to quantify HOIs respectively for the network as a whole, for each link, and for each node. The integration of these measures into the conventional network representation results in a tool for the representation of HOIs \textit{as networks}, which is defined formally using measures of information dynamics, implemented in its linear version by using vector regression models and statistical validation techniques, illustrated in simulated network systems, and finally applied to an illustrative example in the field of network physiology.

\end{abstract}

\maketitle

\section {Introduction}

Complex systems consisting of many interacting units are commonly depicted as networks, according to a paradigm widely used to investigate the structure and dynamics of several real-world phenomena \cite{boccaletti2006complex}. The classical network representation of dynamic systems makes use of nodes, associated to system units and possibly representing temporal dependencies of the unit activity, and links, representing functional dependencies between pairs of units. This representation encoding so-called lower-order interactions (LOIs) is ubiquitous in many fields of science and engineering, as it applies to social systems \cite{benson2016higher}, electronic oscillators \cite{minati2018apparent}, ecosystems and climate systems \cite{woodward2010ecological}, as well as neuroscience and physiology \cite{bashan2012network,bassett2017network}. However, it has now firmly established that in all these systems interactions can occur in groups of three or more nodes, giving rise to collective dynamics known as higher-order interactions (HOIs) that cannot be fully accounted by single-node and pairwise dynamic measures \cite{battiston2020networks}. 

HOIs manifest themselves in complex network systems both at the level of \textit{mechanisms}, through the presence of higher-order terms in generative models of the dynamics at each node, and at the level of \textit{behaviors}, through the emergence of higher-order correlations in the multivariate dynamic processes mapping the system evolution
\cite{rosas2022disentangling}.
Since the dynamics of networks with HOI mechanisms differ substantially to those generated solely by LOI mechanisms \cite{iacopini2019simplicial,gambuzza2021stability}, it is evident that higher-order mechanisms shape in a significant way the behavior of network systems. Nevertheless, higher-order behaviors can emerge even in systems without higher-order mechanisms \cite{rosas2022disentangling}. It is therefore imperative, to fully characterize the dynamic behavior of complex network systems, to move from the use of standard LOI measures to the adoption of frameworks able to depict the various types of HOIs that emerge within the resulting multivariate statistics.

The detection of HOIs from the behavior of network systems embodies the statistical concepts of synergy and redundancy, and is typically formalized into the frame of partial information decomposition and its various developments \cite{williams2010nonnegative,lizier2018information}. Information-theoretic tools play a main role in this context, with the mutual information (MI) being the primary measure to capture pairwise interactions and its multivariate extensions serving as measures of higher-order behaviors. Among the latter, the long-known measure of interaction information (II) \cite{mcgill1954multivariate}, quantifies HOIs in terms of the synergy/redundancy balance among three random variables. The II has been recently extended to an arbitrary number of variables through the concept of O-information (OI) \cite{rosas2019quantifying,stramaglia2021quantifying}. The MI, II, and OI are effective for describing static networks represented by random variables, but they fall short in capturing the dynamics of networks that evolve over time, which typically exhibit autocorrelations and time-lagged cross-correlations. This limitation has been addressed by the recent introduction of information rates for the analysis of higher-order behaviors \cite{faes2022new,sparacino2024measuring}; the resulting measures, i.e. the MI rate (MIR), II rate (IIR), and OI rate (OIR), extend the analysis of HOIs to dynamic networks characterized by random processes.

In this context, the aim of the present study is to introduce a framework for the comprehensive description of different types of high-order behaviors in network systems mapped by multivariate time series. The framework is grounded on the OIR and related measures, which are formulated in this work to assess HOIs operating simultaneously across different levels of resolution: the entire network, individual links, and specific nodes. We use the OIR as a global, network-wise measure capturing HOIs among all the analyzed processes. Further, as the OIR cannot consider higher-order effects that are specific to certain parts of the network (e.g., individual nodes or links), we introduce into the framework two additional HOI descriptors: the local OIR, quantifying the net information shared between two processes and the rest of the network \cite{mijatovic2024assessing}, here taken as link-wise measure of HOIs; and the OIR-gradient \cite{scagliarini2023gradients}, quantifying the information shared between one process and the rest of the network. This new framework allows to capture simultaneously different types of high-order behaviors and to represent them as networks investigated across different levels of resolution.

The rest of this paper is organized as follows. First, we provide the theoretical formulation of OIR, local OIR and OIR-gradient based on combining multiple instances of MIR (Sect. \ref{sec_methods}), and derive a data-efficient implementation of all measures based on linear vector autoregressive (VAR) models (Sect. \ref{prac_comp}), as well as, methods to estimate the measures from multivariate time series (Sect. \ref{estim}) and to assess their statistical significance (Sect. \ref{ss}).
Then, the framework is illustrated in a benchmark simulated network system, depicting the diverse types of high-order behaviors produced even by multivariate processes featuring linear LOI mechanisms (Sect. \ref{valid}).
Finally, the framework is employed to investigate both LOIs and HOIs in an application very common in the fields of computational and network physiology \cite{bashan2012network,cohen2002short,javorka2017basic}, i.e. the study of heart rate, arterial pressure and peripheral vascular resistance time series measured on a beat-by-beat basis from multiple biosignals (Sect. \ref{Appl}); the resulting cardiovascular networks are here investigated in healthy subjects monitored in a resting state and during postural and mental stress, to showcase the relevance of HOIs in the homeostatic regulation analyzed across different physiological states.

\section {Methods}

\subsection {Framework to Assess Higher-Order Interactions at Different Levels of Resolution}
\label{sec_methods}

Let us consider a dynamic network system $\mathcal{X}$ composed of $N$ nodes, 
whose activity is described by $N$ discrete-time random processes $X=\{X_1,\ldots,X_{N}\}$. 
The average rate of information produced over time by the process $X_i$, which maps the activity of the $i^{th}$ node of the network, is quantified by its entropy rate (ER). The ER captures the internal dynamics of the process, i.e., its self-interactions of order one. For a stationary process $X_i$ the ER is defined as \cite{cover1999elements}:
\begin{equation}
 H_{X_i}=\lim_{k \to \infty} \dfrac{1}{k}H(X_{i,n-k:n-1}) = H(X_{i,n}|X^-_{i,n}),
   \label{ER}
\end{equation}
where $X_{i,n}$,  $X_{i,n-k:n-1}$, and $X^-_{i,n}=\lim_{k \to \infty}X_{i,n-k:n-1}$ denote the random variables that sample $X_i$ at the present time $n$, over the past $k$ lags, and over the whole past history, respectively. The second equality in Eq. (\ref{ER}) expresses the ER in the form of conditional entropy, evidencing the amount of information contained in the present time of the process $X_i$ that cannot be explained by its past history: if $X_i$ is a fully random process, it produces information at the maximum rate, resulting in the maximum ER; if, on the contrary, $X_i$ is a fully predictable process, it does not produce new information and its ER is zero.

The interactions of order two between the $i^{th}$  and $j^{th}$  nodes of the analyzed dynamic network can be assessed by the mutual information rate (MIR) between the processes $X_i$ and $X_j$. The MIR  quantifies the information shared between the two processes per unit of time \cite{duncan1970calculation}:
\begin{align}
   I_{X_i;X_j}   &= \lim_{k \to \infty}\frac{1}{k} I(X_{i,n-k:n-1};X_{j,n-k:n-1}), \\ 
   & = H_{X_i}+H_{X_j}-H_{X_i,X_j},
   \label{MIR}
\end{align}
where $H_{X_i}$ and $H_{X_j}$ are the ERs of the processes $X_i$ and $X_j$, and $H_{X_i,X_j}$ is the joint ER of the two processes.

The ER and MIR measures serve as building blocks for computing measures of higher-order interactions (HOIs). In this work we introduce a comprehensive framework for the evaluation of different types of interactions beyond pairwise connections, considering different measures designed to assess HOIs in random variables \cite{rosas2019quantifying} and subsequently refined to quantify dynamic HOIs in random processes \cite{faes2022new}. The framework is grounded on the so-called O-information rate (OIR), a measure capturing the balance between redundant and synergistic HOIs in network of random processes \cite{faes2022new}.
Specifically, the OIR ($\Omega$) of the $N$ random processes grouped in $X$ can be defined elaborating the entropy rates of the processes as:
\begin{equation}
   \Omega_{X}=(N-2)H_X + \sum_{j=1}^{N}[H_{X_j}-H_{X^j}],
   \label{OIR}
\end{equation}
\noindent{where $X^j = X\backslash \{X_j\}$ is a vector process including the activity of all network nodes except $X_j$.} The OIR quantifies collective interactions among all the analyzed processes, and can be also computed in an iterative way from the OIR of a subset including $N-1$ processes, e.g., $X^{j}= X \backslash X_j$, summing a gradient ($\Delta$) which quantifies the informational increment obtained when $X_j$ is added to $X^{j}$ \cite{faes2022new}:
\begin{align}
   \Omega_{X} &=\Omega_{X^{j}} + \Delta_{X_j;X^{j}}, \label{OIR_it} \\
   \Delta_{X_j;X^{j}} &=\sum_{\substack{i=1 \\ i \neq j}}^{N} I_{X_j;X^{ij}}+(2-N)I_{X_j;X^{j}}.
   \label{DeltaOIR}
\end{align}
Note that the OIR is zero if $N=2$, while for $N=3$ it is equal to the OIR-gradient, which in turn corresponds to the so-called interaction information rate (IIR), i.e. $\Omega_{X}=\Delta_{X_j;X^{j}}=I_{X_i;X_j;X_k}= I_{X_i;X_j}+I_{X_k;X_j}-I_{X_i,X_k;X_j}$ ($i\neq j\neq k \in\{1,2,3\}$) \cite{faes2022new}. Then, when networks mapped by more than three random processes are analyzed, the IIR can be generalized by focusing on two processes $X_i$ and $X_j$ and collecting all other processes in the vector $X^{ij} = X\backslash \{X_i, X_j\}$; the resulting measure is the dynamic version of the local OI defined in \cite{rosas2019quantifying}, also denoted as net information shared in \cite{mijatovic2024assessing}:
\begin{equation}
I_{X_i;X_j;X^{ij}} = I_{X_i;X_j} + I_{X_i;X^{ij}} - I_{X_i;X^{i}},
\label{IIR}
\end{equation}
\noindent{where $X^{ij} = X\backslash \{X_i, X_j\}$ include all network nodes except $X_i$ and $X_j$.}

Importantly, all the measures defined in Eqs. (\ref{OIR}), (\ref{DeltaOIR}) and (\ref{IIR}) quantify HOIs by capturing the balance between \textit{redundancy} and \textit{synergy}, taking positive values when redundant interactions prevail over synergistic interactions, and negative values in the opposite case. Here, the terms "redundant" and "synergistic" applied to multivariate information measures refer to information identically provided by each of the multiple analyzed units, and to new information that emerges when the units are considered together.
Building on these measures, our framework allows exploring HOIs within the observed network system simultaneously across multiple levels of spatial resolution. This includes:

\begin{itemize}

\item \textit{Node-wise} analysis of HOIs based on the OIR-gradient  (Eq. \ref{DeltaOIR}), quantifying the informational character (redundant or synergistic) provided by the inclusion of the analyzed node into the multiplet formed by all other network nodes;

\item \textit{Link-wise} analysis of HOIs based on the local OIR (Eq. \ref{IIR}), quantifying the net information shared between the two analyzed nodes and the rest of the network;

\item \textit{Network-wise} analysis of HOIs based on the OIR (Eq. \ref{OIR}), quantifying the redundant/synergistic character of the overall dynamic interactions among all the nodes of the analyzed network.

\end{itemize}


\subsection {Computation for Linear Systems}
\label{prac_comp}

A common and convenient approach to describe the dynamics
of the zero-mean vector process $X$ is to use the linear vector autoregressive (VAR) model \cite{lutkepohl2005new}:
\begin{equation}
X_n=\sum_{k=1}^{p}\textbf{A}_kX_{n-k}+U_n,
\label{fullVAR}
\end{equation}
where $p$ is the model order, $X_n$ and $X_{n-k}$ are the $N-$dimensional variables sampling the process $X$ at the present time time $n$ and at $k$ lags in the past, $\textbf{A}_k$ are $N \times N$ coefficient 
matrices encoding the causal relations from past to present within and between processes for each lag $k$, and $U$ is an i.i.d. innovation process with $N \times N$ 
covariance matrix $\mathbf{\Sigma}_U=\mathbb{E}[U_nU_n^{\intercal}]$.

Starting from the ubiquitous representation in Eq. (\ref{fullVAR}), developments of the theory of Granger causality have shown that it is possible to define \textit{restricted} models which describe the dynamics of properly chosen subsets of processes \cite{faes2017information}.
Specifically, the dynamics of the generic sub-process $Y \subset X$ collecting $M$ scalar processes taken from $X$ (typically, $M<N$) are described by the restricted model:
\begin{equation}
Y_n=\sum_{k=1}^{q}\textbf{B}_kY_{n-k}+W_{Y,n},
\label{restrictedVAR}
\end{equation}
\noindent{where $W_Y$ is the $M$-dimensional innovation process feeding the restricted model to yield the analyzed sub-process $Y$, and $\textbf{B}_k$ are the relevant $M \times M$ coefficient matrices; note that the order of the restricted model $q$ will generally tend to infinity because a sub-process of a VAR process will have a moving average component \cite{barnett2015granger}.} 
Importantly, under the assumption of joint Gaussianity for the overall process $X$, the ER of the sub-process $Y$ can be derived straightly based on the generalized variance of the residuals of the restricted model  of Eq. (\ref{restrictedVAR}) as follows \cite{sparacino2024measuring}:
\begin{equation}
    H_{Y}=\frac{1}{2} \log (2\pi e)^M |\mathbf{\Sigma}_{W_Y}|,
   \label{ERrestrictedVAR}
\end{equation}
where $\mathbf{\Sigma}_{W_Y}=\mathbb{E}[W_{Y,n}W_{Y,n}^{\intercal}]$ is the $M \times M$ covariance matrix of the restricted innovation process $W_Y$. Moreover, considering another sub-process $Z \subset X$ disjoint from $Y$, and considering the joint process $[YZ]=Y\cup Z$, similar formulations like in Eqs. (\ref{restrictedVAR}) and (\ref{ERrestrictedVAR}) can be intuitively written and combined to derive the MIR between $Y$ and $Z$ as follows:
\begin{equation}
    I_{Y;Z}=\frac{1}{2} \log \frac{|\mathbf{\Sigma}_{W_Y}| |\mathbf{\Sigma}_{W_Z}|}{|\mathbf{\Sigma}_{W_{YZ}}|}.
   \label{MIRrestrictedVAR}
\end{equation}

Eqs. (\ref{ERrestrictedVAR}) and (\ref{MIRrestrictedVAR}) are the cornerstone for the computation of any information dynamic measure based on VAR models. In fact, the ER of the scalar process $X_i$ is obtained simply by assuming $Y=X_i$ in Eq. (\ref{ERrestrictedVAR}):
\begin{equation}
H_{X_i}=\frac{1}{2} \log 2\pi e \Sigma_{W_{X_i}};
\label{ERgauss}
\end{equation}
the MIR between the processes $X_i$ and $X_j$ is obtained simply by assuming $Y=X_i$ and $Z=X_j$ in Eq. (\ref{MIRrestrictedVAR}):
\begin{equation}
I_{X_i;X_j}=\frac{1}{2} \log \frac{\Sigma_{W_{X_i}} \Sigma_{W_{X_j}}}{|\mathbf{\Sigma}_{W_{X_iX_j}}|};
\label{MIRgauss}
\end{equation}
and the local OIR between $X_i$ and $X_j$ given the rest of the network is derived applying Eq. (\ref{MIRgauss}) three times according to Eq. (\ref{IIR}) to yield:
\begin{equation}
I_{X_i;X_j;X^{ij}}=\frac{1}{2} \log \frac{\Sigma_{W_{X_i}} \Sigma_{W_{X_j}} |\mathbf{\Sigma}_{W_{X^{ij}}}||\mathbf{\Sigma}_{U}| } {|\mathbf{\Sigma}_{W_{X_{ij}}}||\mathbf{\Sigma}_{W_{X^j}}||\mathbf{\Sigma}_{W_{X^i}}|}.
\label{IIRgauss}
\end{equation}
With similar reasoning, the OIR among all the processes in $X$ can be derived from Eq. (\ref{OIR}) using the formulation of the ER (Eq. \ref{ERrestrictedVAR}), one time with $Y=X$ and repeatedly with $Y=X_j$ and $Y=X^j$, and the OIR-gradient of the process $X_j$ given the other processes $X^j$ can be derived from Eq. (\ref{DeltaOIR}) using the formulation of the MIR ( Eq. (\ref{MIRrestrictedVAR}), one time with $\{Y=X_j,Z=X^j \}$ and repeatedly with $\{Y=X_j,Z=X^{ij} \}$.

\subsection {Estimation}
\label{estim}

The parameters of the restricted model (Eq. \ref{restrictedVAR}), i.e. the coefficients $\mathbf{B}_k$ and the covariance of the residuals $\mathbf{\Sigma}_{W_Y}$, can be derived from the parameters of the full model (Eq. \ref{fullVAR}), $\textbf{A}_k$ and $\mathbf{\Sigma}_U$, through a two-step procedure which (i) derives the time-lagged covariance structure of the full process $X$, and (ii) reorganizes such structure to relate it to the covariance of the sub-process $Y$ \cite{barnett2014mvgc}.

The step (i) exploits the fact that the covariance of $X$, $\mathbf{\Sigma}_{X, k}=\mathbb{E}[X_nX_{n-k}^{\intercal}]$,  is related to the VAR parameters via the well-known Yule-Walker equations \cite{lutkepohl2005new}:
\begin{equation}
    \mathbf{\Sigma}_{X, k} = \sum_{l=1}^{p} \mathbf{A}_l \mathbf{\Sigma}_{X, k-l} + \delta_{k0}\mathbf{\Sigma}_U,
    \label{YW}
\end{equation}
where $\delta_{k0}$ is the Kronecher delta function. To solve this equation for $k=0, 1, \dots, p-1$, we first express Eq. (\ref{fullVAR}) in a compact form as $X^p_n = \mathbf{A}^pX^p_{n-1} + U^p_n$, where
\begin{subequations}
\begin{align}
X_n^p &= \begin{bmatrix}
X_{n}^{\intercal} X_{n-1}^{\intercal}\cdots X_{n-p+1}^{\intercal}
\end{bmatrix}^{\intercal}, \\
\mathbf{A}^p &= \begin{bmatrix} 
   \mathbf{A}_1 & \cdots & \mathbf{A}_{p-1} & \mathbf{A}_p \\
    \mathbf{I}& \cdots & \mathbf{0} & \mathbf{0} \\
     \vdots & \ddots & \vdots  & \vdots\\
       \mathbf{0}& \cdots & \mathbf{I} & \mathbf{0}\end{bmatrix}, \\
U^p_n &= \begin{bmatrix} U_n^{\intercal}  \mathbf{0}
\end{bmatrix}^{\intercal}.
\end{align}
\label{fullVar_cf}
\end{subequations}
Then, the covariance matrix of $X^p_n$, $\mathbf{\Sigma}^p_X = \mathbb{E} [X^p_n {X_n^p}^\intercal]$, which has the form:
\begin{equation}
\mathbf{\Sigma}^p_X = \begin{bmatrix} 
   \mathbf{\Sigma}_X &  \mathbf{\Sigma}_{X,1} & \cdots & \mathbf{\Sigma}_{X, p-1}\\
   \mathbf{\Sigma}_{X,1}^{\intercal} &  \mathbf{\Sigma}_X & \cdots & \mathbf{\Sigma}_{X, p-2} \\
    \vdots & \vdots & \ddots  & \vdots\\
       \mathbf{\Sigma}_{X, p-1}^{\intercal} &  \mathbf{\Sigma}_{X, p-2}^{\intercal} & \cdots & \mathbf{\Sigma}_X \\
      \end{bmatrix},
       \label{cov_mat}
\end{equation}
can be expressed as:
\begin{equation}
\mathbf{\Sigma}_X^p =\mathbf{A}^p\mathbf{\Sigma}_X^p(\mathbf{A}^p)^{\intercal} + \mathbf{\Sigma}_\textit{U}^p,
\label{Lyapunov}
\end{equation}
which is a discrete-time Lyapunov equation ($\mathbf{\Sigma}_\textit{U}^p$ denotes the covariance of $U_n^p$, i.e., 
$\mathbf{\Sigma}^p_U = \mathbb{E} [U^p_n {U_n^p}^\intercal]$). The Lyapunov equation
can be solved for $\mathbf{\Sigma}_X^p$, thus yielding the covariance matrices $\mathbf{\Sigma}_X, \mathbf{\Sigma}_{X,1}, \cdots, \mathbf{\Sigma}_{X, p-1}$. Then, the covariance
can be calculated recursively for any lag $k \geq p$  by applying Eq. (\ref{YW}), so as to obtain $\mathbf{\Sigma}_{X,p}, \mathbf{\Sigma}_{X,p+1}, \cdots, \mathbf{\Sigma}_{X, q}$. 

The step (ii) of the estimation procedure starts with extracting, from the covariance matrices $\mathbf{\Sigma}_{X, k}$ computed for each lag $k \in \{0,1,\ldots,q\}$, only the covariance elements relevant to the sub-process $Y$, which form the covariance matrices $\mathbf{\Sigma}_{Y, k}=\mathbb{E}[Y_nY_{n-k}^{\intercal}]$. These covariances are arranged to obtain the covariance matrix of $Y$, $\mathbf{\Sigma}_Y =  \mathbb{E} [Y_n Y_n^\intercal]$, the covariance of the predictors of Eq. (\ref{restrictedVAR}), $\mathbf{\Sigma}_Y^q = \mathbb{E} [Y_n^q {Y_n^q}^\intercal]$, and the cross-covariance $\mathbf{\Sigma}_{Y_n; Y_{n-1}^q} = \mathbb{E} [Y_n Y_{n-1}^{q^\intercal}]$ (where $Y_{n-1}^q = \begin{bmatrix} Y_{n-1}
^{\intercal} \cdots Y_{n-q}^{\intercal} \end{bmatrix}^{\intercal}$). Finally, solving the Yule-Walker equations, the coefficients of the restricted model are computed as $\mathbf{B}_k=\mathbf{\Sigma}_{Y_n; Y_n^q} {\mathbf{\Sigma}_{Y}^q}^{-1}$, and the covariance matrix of the residuals is obtained as \cite{barnett2014mvgc}:
\begin{equation}
\mathbf{\Sigma}_{W_Y}= \mathbf{\Sigma}_{Y} - \mathbf{\Sigma}_{Y_n; Y_n^q} {\mathbf{\Sigma}_{Y}^q}^{-\intercal} \mathbf{\Sigma}_{Y_n; Y_n^q}^\intercal.
\label{sigma_res}
\end{equation}

The procedure described above starts from the VAR parameters $\textbf{A}_k$ and $\mathbf{\Sigma}_U$ of the full model, which can be easily estimated from realizations of the process $X$ available in the form of multivariate time series. While advanced procedures exist to perform VAR model identification, e.g. using penalized regression to derive sparse models in high-dimensional settings \cite{antonacci2020information} or using time-varying regression to derive time-dependent parameters in non-stationary conditions \cite{antonacci2023time},  the classical least squares approach was followed in this work \cite{lutkepohl2005new}.
The order $p$ of the full model is typically estimated using information-theoretic criteria, like the Akaike  \cite{akaike1974new}, or the Bayesian  \cite{schwarz1978estimating}, while the order $q$ of the restricted models is set at high values to capture the decay of the correlations at increasing lags \cite{barnett2014mvgc,faes2017information}.


\subsection {Statistical Significance}
\label{ss}

This section describes the approach followed to statistically validate measures of the OIR-gradient, local OIR, and OIR, i.e., to determine if an estimated value of any of these measures can be regarded as significantly different from zero, detecting the presence of redundancy if positive and of synergy if negative. In this work, a procedure based on the bootstrap technique \cite{efron1979bootstrap_first} was implemented, exploiting the fact that the HOI measures can take both positive and negative values. Specifically, for any given multivariate time series, a VAR model was identified by the least squares method, and bootstrap pseudo-series were created by feeding the identified model with new realizations of the innovations $U$ obtained by repeatedly sampling with replacement from the estimated innovations. This procedure ensures that the bootstrap series retain all characteristics of the original process $X$, including causal relationships and zero-lag correlations.
Then, the OIR-gradient, local OIR and OIR are computed from the bootstrap  realizations. The procedure is repeated several times to construct bootstrap distributions of each HOI measure, whose confidence intervals are finally used to assess the statistical significance of each original estimate of the measure. Setting a significance level $\alpha$, if the zero level falls between the $(100\cdot\alpha/2)^{th}$ and $(100\cdot(1-\alpha/2))^{th}$ percentiles of the bootstrap distribution of the HOI measure, the measure itself is considered as non-significant; otherwise, it is deemed as significant \cite{sparacino2023statistical}.

As regards the detection of the statistical significance of the MIR estimates, the bootstrap technique cannot be followed because any MIR estimate takes only positive values. In this case, the method of surrogate data was employed, generating iterative amplitude-adjusted Fourier Transform surrogates \cite{schreiber1996improved} of each set of original time series. While preserving the individual properties (amplitude distribution, autocorrelation structure) of each original time series, these surrogates are fully uncoupled, thus adhering to the null hypothesis of uncorrelated time series for which the MIR is absent. The rejection of this null hypothesis occurs with significance $\alpha$ when the MIR estimated on the original series exceeds the $(100\cdot(1-\alpha))^{th}$ percentile of the surrogate distribution.

\section{Validation on Simulations}
\label{valid}

In this section we illustrate the framework described above in a simulated dynamic network whose activity is mapped by a multivariate random process described by a VAR model. First, the HOIs measures (local OIR, OIR and OIR-gradient) are computed analytically from the theoretical values imposed for the model parameters. Then, all measures are estimated from finite-length realizations of the network process, and their statistical significance is assessed at varying the time series length.

\subsection {Theoretical Analysis}
\label{theor_ex}

\begin{figure*} [t!]
\centering
\includegraphics[scale=0.93]{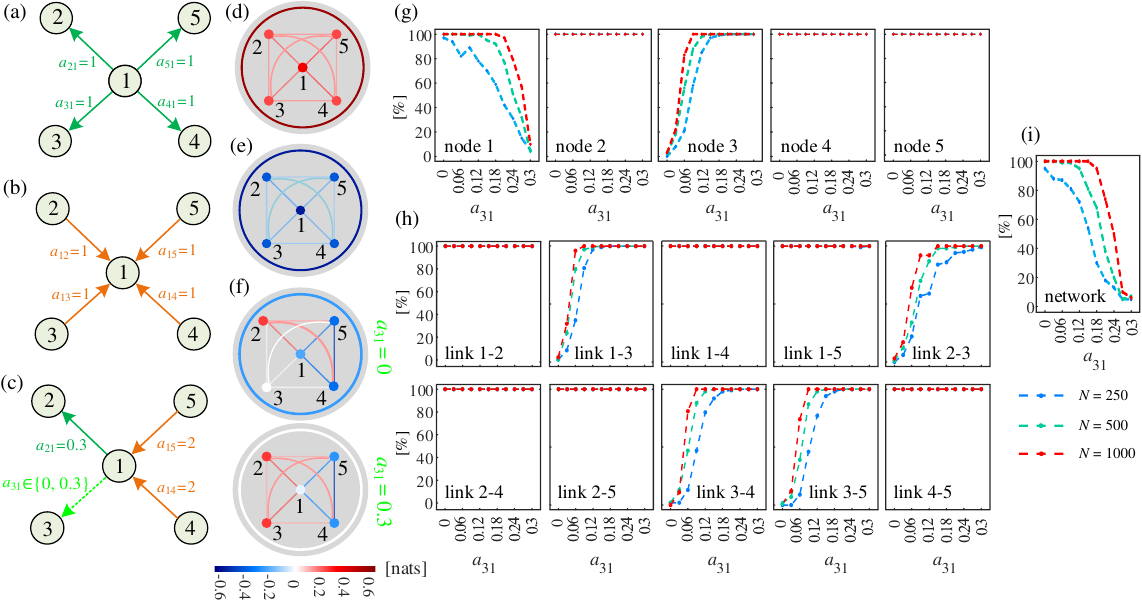}
\caption{Analysis of a simulated network of random processes connected via a hub (node 1) communicating with four leaves (nodes 2-5), where the hub acts: (a) as a \textit{source} of information sent to all leaves (green arrows); (b) as a \textit{sink} of information received from all leaves (orange arrows); (c) as a \textit{mediator} of the information received from nodes 4-5 and sent to nodes 2-3.
The corresponding higher-order networks computed from the VAR parameters by measuring the theoretical values of the synergy/redundancy balance for each node, each link and the whole network, respectively using the OIR-gradient (Eq. \ref{DeltaOIR}), the local OIR (Eq. \ref{IIR}), and the OIR (Eq. \ref{OIR}), are shown for the configurations with the hub acting as source (d), sink (e), and mediator (f, where the cases with node 3 isolated or connected to the network are distinguished); each measure is represented using pseudo-colors where negative values (synergy), values close to zero (balance) and positive values (redundancy) are denoted respectively by shades of blue, white and red. When the same measures are computed from realizations of the network configuration in (c) varying the time series length $N$ and the coupling coefficient $a_{31}$, the percentage over 100 realizations of statistically significant estimates detected via the bootstrap data analysis are shown in (g) for the OIR-gradient of each node, in (h) for the local OIR of each link, and in (i) for the OIR of the whole network.}
\label{simu_res}
\end{figure*}

We consider $N=5$ processes interacting in a \textit{star} structure as depicted  in Fig. \ref{simu_res}a-c, where node 1 and nodes 2-5 act respectively as the hub and as leaves. The processes are described by a VAR model fed by independent Gaussian innovations with zero mean and unit variance, formulated as in Eq. (\ref{fullVAR}) with order $p=1$. We analyze three configurations with different interactions all occurring at lag 1, obtained imposing specific values of the coefficients $a_{ij}$ of the matrix $\textbf{A}_1$. In the three cases, the hub acts: (i) as a \textit{source}, sending information to all leaves (Fig. \ref{simu_res}a); (ii) as a \textit{sink}, receiving information from all leaves (Fig. \ref{simu_res}b); and (iii) both as \textit{source} and as  \textit{sink}, sending to nodes 2-3 the information that is received from nodes 4-5 (Fig. \ref{simu_res}c).

The HOIs networks built from the theoretical values of the VAR parameters in the three cases are shown in Fig. \ref{simu_res}d-f, reporting the color-coded values (blue shades: synergy; red shades, redundancy; white: balance) of the OIR-gradient (color of each node), of the local OIR (color of each link), and of the OIR (color of the external circle). The use of exact values of the three measures in the different configurations allows to illustrate their capability to capture the informational character of the HOIs among hubs and leaves at different levels of spatial resolution. Specifically, when the hub is a source of information for all leaves, all measures reveal the predominance of redundant interactions (Fig. \ref{simu_res}d), induced by the simulation of multiple common drive effects (Fig. \ref{simu_res}a). When the hub is a sink of information arriving from the leaves, all measures reveal the synergistic character of HOIs at all levels of resolution (Fig. \ref{simu_res}e), arising from the presence of multiple common child effects (Fig. \ref{simu_res}b). 
In the third configuration with the hub acting as mediator, the presence of common driver, cascade and common child effects determines an heterogeneous distribution of HOIs, with nodes 4-5 and their connected links denoting synergy and node 2 and its connected links denoting redundancy (Fig. \ref{simu_res}f). The two settings relevant to node 3 evidence further interesting behaviors: when node 3 is isolated ($a_{31}=0$), its contribution to HOIs vanishes for both the OIR-gradient and the local OIR (white node and links in Fig. \ref{simu_res}f, upper panel); when node 3 receives information from node 1 ($a_{31}=0.3$), it brings redundant HOIs to the network in terms of OIR-gradient and local OIR (red node and links in Fig. \ref{simu_res}f, lower panel). Interestingly, the global OIR captures the net contributions locally by the HOIs: the overall interaction among the five nodes is mostly synergistic (blue external circle) when node 3 is disconnected, and is balanced (white external circle) when node 3 behaves like node 2 and in opposition with nodes 4,5.

\subsection{Analysis on Simulated Time Series}

Here we report the practical analysis of the network process defined in Sect. \ref{theor_ex}, realized in the third configuration where the hub act as mediator of information. The process was implemented by feeding the VAR model of Eq. (\ref{fullVAR}) ($p=1$) with realizations of five independent Gaussian white noises with unit variance, setting the time-lagged coefficients as in Fig. \ref{simu_res}c. Specifically, the coefficient $a_{31}$ was varied between 0 and 0.3 in steps of 0.03, and for each setting the simulation was run 100 times changing the length of the generated time series in the range $N=\{250, 500, 1000\}$. For each realization, we estimated the measures of the OIR-gradient, local OIR, and OIR from the VAR parameters estimated through least squares identification as described in Sect. \ref{estim} (order of the full model: $p=1$; order of the restricted models: $q=20$ \cite{faes2017information}).
Then, the statistical significance of each HOI measure was assessed individually for each simulated time series using the bootstrap method described in Sect. \ref{ss}, implemented over 100 surrogates with significance level $\alpha=0.05$.

The HOI measures were estimated with minimal bias (OIR-gradient: 0.03 for $N=250$, 0.02 for $N=500$, 0.01 for $N=1000$; local OIR: 0.02 for $N=250$, 0.01 for $N=500$, 0.01 for $N=1000$; OIR: 0.05 for $N=250$, 0.03 for $N=500$, 0.02 for $N=1000$; global average across nodes, links and conditions) and with standard deviation decreasing with the time series length (OIR-gradient: 0.04 for $N=250$, 0.03 for $N=500$, 0.02 for $N=1000$; local OIR: 0.03 for $N=250$, 0.02 for $N=500$, 0.01 for $N=1000$; OIR: 0.07 for $N=250$, 0.04 for $N=500$, 0.03 for $N=1000$; global average across nodes, links and conditions).

The percentage of realizations for which the HOI measures were detected as statistically significant at varying the coupling coefficient $a_{31}$ is reported in Fig. \ref{simu_res}g for the OIR-gradient computed across nodes, in Fig. \ref{simu_res}h for the local OIR computed across links, and in Fig. \ref{simu_res}i for the OIR computed for the whole network.
The variation of the coefficient $a_{31}$ leads to a gradual transition of the network structure from isolation of node 3 to its connection to node 1, which is associated with the rise of node-wise and link-wise HOIs involving node 3, to the dampening of synergistic node-wise HOIs involving node 1, and to the shift from synergistic to balanced network-wise HOIs (see Fig. \ref{simu_res}c, f). These transitions were empirically supported by the progressive increase of the percentage of OIR-gradient and local OIR estimates involving node 3 that were identified as statistically significant, rising from the nominal 5 \% to 100 \% (Fig. \ref{simu_res}g, h), as well as by the progressive decrease of the percentage of statistically significant global OIR estimates, decreasing from 100\% to 5\% (Fig. \ref{simu_res}i); these variations were detected more rapidly for longer time series. On the contrary, the node-wise and link-wise HOIs not involving node 3 remained unaffected by these transitions, and their redundant or sinergistic nature was detected with 100\% sensitivity for all time series lengths.

\section {Application to Cardiovascular Dynamic Networks}
\label{Appl}

This section presents the application of the proposed framework to cardiovascular networks investigated under different physiological conditions. Specifically, we explore the physiological network related to the short-term neural control of cardiovascular oscillations in the resting state and during orthostatic and mental stress, analyzing beat-to-beat variability series representing heart rate, arterial pressure, cardiac output and peripheral vascular resistance \cite{krohova2020vascular}. We first describe lower-order interactions (LOIs) analyzing node activity and pairwise connectivity through the ER and MIR measures, and then move to HOIs described at different levels of resolution via the OIR, the local OIR and the OIR-gradient. Our focus is on investigating how different types of HOIs emerge from lower-order dynamics in cardiovascular networks across different physiological states.

\subsection {Data and Experimental Protocol}

We re-analyze a dataset previously collected to study the effects of postural stress and mental stress on cardiovascular regulation \cite{krohova2020vascular}. While the original work encompassed larger groups of healthy subjects, here we consider a subset of the 26 participants for whom the physiological time series of interest were successfully collected in all three physiological conditions.

Specifically, five physiological time series were measured on a beat-to-beat basis from the electrocardiogram (ECG), finger arterial blood pressure (ABP), and impedance cardiography (ICG) signals, recorded simultaneously and digitized at 1 KHz in the resting supine position (R), during postural stress induced by passive head-up tilt at a 45$^{\circ}$ angle (T), and during mental stress induced by performing arithmetic tests in the supine position (M). 
The five analyzed series, constituting realizations of the processes $X=\{X_1,\ldots,X_5\}$ describing the cardiovascular network, were the heart period (HP, process $X_1=H$), the systolic pressure (SP, process $X_2=S$), the diastolic pressure (DP, process $X_3=D$), the cardiac output (CO, process $X_4=C$), and the peripheral vascular resistance (PR, process $X_5=P$). These series were derived according to a well-established measurement convention \cite{javorka2017basic,krohova2020vascular,faes2017information}. In detail, the $n^{th}$ sample ($n=1,\ldots,300$) of each series was obtained as follows: $H_n$ was taken as the duration of the interval between two consecutive R peaks of the ECG signal; $S_n$ was taken as the maximum value of the ABP signal measured inside $H_n$; $D_n$ was taken as the minimum value of the ABP signal measured between the occurrence times of $S_n$ and $S_{n+1}$; $C_n= 60 \cdot (V_n/H_{n-1})$, where $V_n$ is  the stroke volume sequence obtained from the ICG signal within $H_n$; and $P_n=M_n/C_n$, where $M_n$ is the mean ABP measured as the average ABP between the occurrence times of $D_{n-1}$ and $D_{n}$. Further details about the protocol and the measurement can be found in \cite{krohova2020vascular,mijatovic2024assessing}.

\subsection {Data Analysis and Results}

For each subject and condition, each of the selected time series was normalized by subtracting the mean and dividing by the standard deviation. The analysis was then performed according to the procedure described in Sects. \ref{prac_comp} and \ref{estim}, optimizing the order $p$ of the full VAR model through the Bayesian criterion \cite{schwarz1978estimating} and using $q$ = 20 lags to identify restricted VAR models \cite{mijatovic2024assessing}. Statistical validation of MIR and OIR was performed according to Sect. \ref{ss}, \textcolor{black}{generating 100 surrogate and bootstrap time series} and setting a significance level $\alpha=0.05$.

\begin{figure} [t!]
\centering
\includegraphics[scale=0.9]{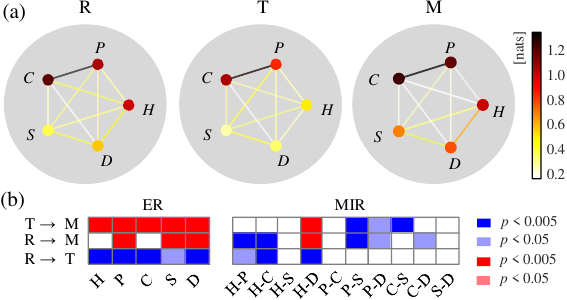}
\caption{Lower-order interactions within and between the $N=5$ processes of heart period $H$, systolic pressure $S$, diastolic pressure $D$, cardiac output $C$, and peripheral resistance $P$. (a) Color-coded maps depicting the median values across subjects of the ER (node color) and of the MIR (link color) computed in the resting state (R) and during stress conditions induced by head-up tilt (T) and mental arithmetics (M). (b) Statistically significant differences (blue boxes, decreasing values; red boxes, increasing values) between median values of the ER and MIR distributions (comparisons: R $\rightarrow$ T, R $\rightarrow$ M, T $\rightarrow$ M), assessed by the Wilcoxon signed-rank test.}
\label{app_cardio_ER_MIR}
\end{figure}

The networks of LOIs obtained computing the ER of each cardiovascular process ($H$, $D$, $S$, $C$, $P$) and the MIR between each pair of processes in each physiological condition (R, T, M) are depicted in Fig. \ref{app_cardio_ER_MIR}a  as color-coded values of nodes and links (\textcolor{black}{median} over 26 subjects), while the statistically significant differences between pairs of conditions are reported in Fig. \ref{app_cardio_ER_MIR}b.
The analysis of the individual node dynamics performed by the ER measure documented a statistically significant decrease from R to T and a significant increase from T to M for each process, as well as significant increase for PR, SP and DP from R to M.
These findings indicate reduced complexity (higher regularity) of all physiological variability series during postural stress compared to rest, which can be ascribed to the expected parasympathetic control inhibition reflected mostly in heart period dynamics and the 
sympathetic nervous activity activation related to postural stress producing
a rise of low-frequency regular oscillations \cite{faes2019comparison,mijatovic2022measuring}. On the other hand, mental stress induced higher complexity of arterial pressure and vascular resistance series also if compared to the resting state, suggesting higher involvement of sympathetic
vascular control and a decreased baroreflex buffering of blood pressure oscillations mostly reflecting reflex parasympathetic control inhibition.


\begin{figure*} [t!]
\centering
\includegraphics[scale=0.9]{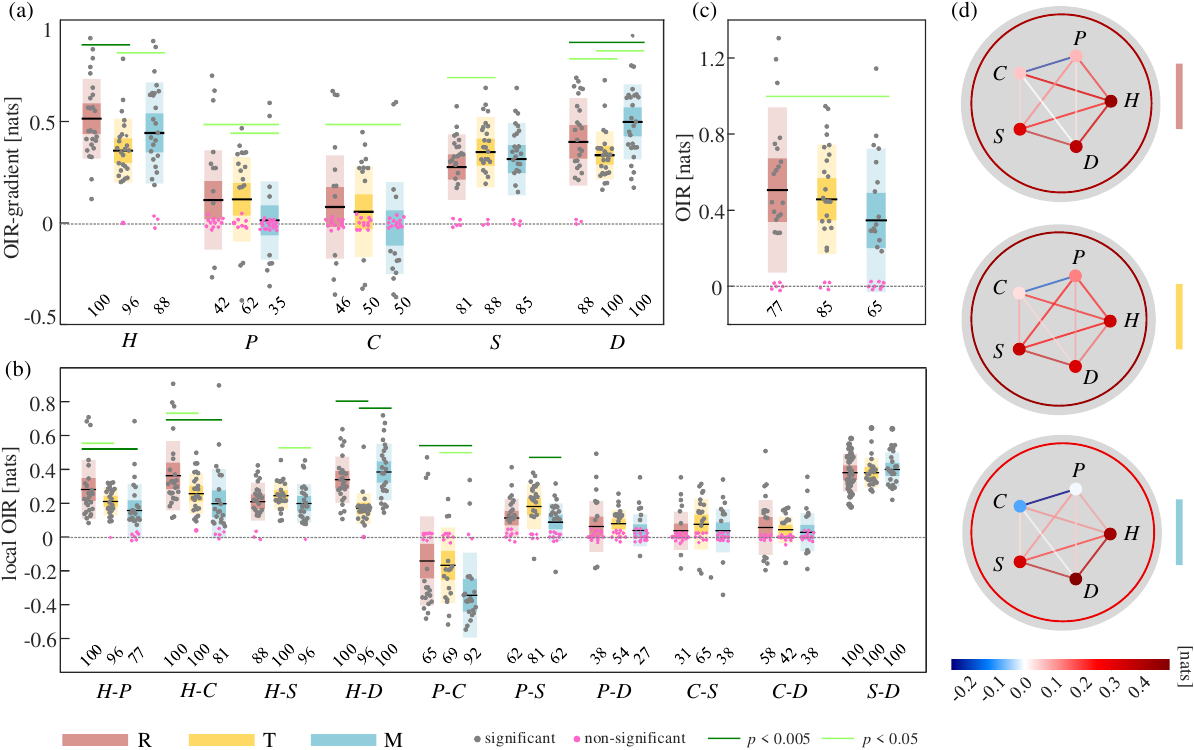}
\caption{Higher-order interactions in the cardiovascular network mapped by the $N=5$ processes of heart period $H$, systolic pressure $S$, diastolic pressure $D$, cardiac output $C$, and peripheral resistance $P$, quantified at the level of nodes by the OIR-gradient (a), of links by the local OIR (b), and of the whole network by the OIR (c), and expressed distributions  across subjects (boxplot and individual values) obtained during resting (R),  head-up tilt (T) and mental arithmetic (M) conditions. The numbers below the distributions indicate the percentage of subjects for whom the estimates of HOIs measures were evaluated as statistically significant through the bootstrap data analysis (gray dots: significant values, purple dots: non-significant); green lines denote statistically significant differences between pairs of distributions (comparisons: R vs. T, R vs. M, T vs. M) assessed with the Wilcoxon signed-rank test. (d) Color-coded maps depicting the median values across subjects of the OIR-gradient (node color), local OIR (link color), and OIR (color of the external circle) computed in the R, T, and M conditions.}
\label{app_cardio_HOis}
\end{figure*}

As regards the MIR, it was always detected as statistically significant using the iAAFT surrogate procedure, revealing a fully connected network of pairwise physiological interactions.
The strongest coupling in the network was detected between PR and CO for each condition, resulting from the tight inverse relation between these two processes \cite{krohova2020vascular}. The MIR was high also for the links SP-HP, HP-DP and SP-DP, reflecting typical mechanisms of cardiovascular coupling like the baroreflex feedback (SP $\rightarrow$ HP) and the mechanical feed-forward (HP $\rightarrow$ SP), the cardiac run-off (HP $\rightarrow$ DP), and the close association between DP and neighboring SP values (DP $\rightarrow$ SP and SP $\rightarrow$ DP) 
\cite{cohen2002short,javorka2017basic, faes2013mechanisms}.

The undirected coupling strength reflected by the MIR for these links was rather stable across conditions, except for the link HP-DP whose strength decreased during T and increased during M;
this suggests that postural stress and mental stress have different impact on the dynamics resulting from the cardiac run-off mechanism   \cite{javorka2017basic}. The remaining links, i.e. those involving the variability of PR or CO together with the other processes, were less strong, though still statistically significant; for these links, the MIR decreased with mental stress even if the ER increased, documenting that stronger coupling can be observed in parallel with weaker regularity of the dynamics \cite{porta2014effect}.

Fig. \ref{app_cardio_HOis} depicts the results of the analysis of HOIs in the analyzed cardiovascular network, reporting the distributions of OIR-gradient (a), local OIR (b) and OIR (c), computed for each node, link and for the whole network in the three conditions, as well as, the HOI networks depicted using color-coded maps of the median values of each measure computed across subjects (d).
The network-wise analysis of HOIs suggests the presence of dominant redundancy in the analyzed cardiovascular network, as documented by the positive values of the OIR measure in all conditions (Fig. \ref{app_cardio_HOis}c). Redundancy in cardiovascular networks can arise as a consequence of the action of common physiological drivers like respiration or sympathetic activity modulating simultaneously several of the analyzed processes \cite{porta2017quantifying,porta2018paced,scagliarini2024gradients,mijatovic2024assessing}.
Nevertheless, the analysis performed with different resolution reveals an heterogeneous distribution of HOIs at the levels of nodes and links. Indeed, Fig. \ref{app_cardio_HOis} a, b show that: (i) the OIR-gradient relevant to the HP dynamics and the local OIR of the related links (i.e., HP-PR, HP-CO, HP-SP, HP-DP), as well as the OIR-gradients of DP and SP and their related local OIR (SP-DP), are consistently redundant and significant for the majority of subjects; (ii) the OIR-gradient of PR and CO and the local OIR of the links PR-SP, PR-DP, CO-SP, CO-DP are more balanced and significant in about half of the subjects; and (iii) the local OIR involving PR and CO displays evident synergy, statistically significant in most subjects. Heterogeneity in the distribution of HOIs across the nodes of the physiological network is an indicator of the coexistence of common drive mechanisms like those related to respiration affecting both heart rate and arterial pressure variability \cite{porta2018paced,porta2021categorizing}, and of common target mechanisms like those associated to the influence of the dynamics of vascular resistance and stroke volume on arterial pressure \cite{nardone2018evidence, mijatovic2024assessing}.

The transition from rest to stress determined a reconfiguration of the networks of HOIs, which show a general tendency towards lower net redundancy or the emergence of net synergy. In particular, we find that postural stress is associated with a marked reduction of the OIR-gradient relevant to HP and DP, as well as of the corresponding local OIR (HP-DP). Similarly, mental stress is associated with a significant reduction of the OIR-gradient relevant to PR and CO, as well as of their local OIR (PR-CO). Moreover, the local OIR of the links HP-PR and HP-CO decreased significantly with both types of stressor. At the network-wise level, the values of net redundancy were comparable at rest and during postural stress, and significantly lower during mental stress (Fig. \ref{app_cardio_HOis}c, d).
Methodologically, a decrease of redundancy and/or an increase in synergy occur when HOI measures like the OIR-gradient or the local OIR are computed for nodes acting as sources of information that is injected in the network when their dynamics are considered (see simulations, Fig. 1e, f). Our results show that this happens for HP and DP during head-up tilt, and for PR and CO during the execution of mental calculations. This suggests that synergistic mechanical effects like the cardiovascular feedforward (HP $\rightarrow$ SAP) \cite{javorka2017causal} are strengthened by postural stress, and synergistic autonomic effects like the sympathetic-driven modulation of the contractility of heart and vessels influencing arterial pressure (effects CO $\rightarrow$ SAP and PR $\rightarrow$ SAP) \cite{thomas2011neural} are strengthened by mental stress.

\section{Conclusions and Perspectives}

This work introduces a framework for evaluating higher-order statistical dependencies in dynamic network systems, explicitly designed to explore different levels of resolution. The framework incorporates dynamic implementations of the O-information, its local counterpart, and its gradient, which are intended to quantify HOIs respectively for the network as a whole, for each link, and for each node.
As such, it allows not only to identify the emergent redundant or synergistic character of a dynamic network system, but also to assess how the synergy/redundancy balance is distributed across links and nodes.
The existing approaches proposed to assess spatially-resolved high-order behaviors, which focus on computing HOI measures on subsets of nodes (multiplets) of the  network at hand \cite{stramaglia2021quantifying,marinazzo2022information}, analyze each given multiplet ignoring the rest of the network, and are impractical to implement as the number of multiplets to analyze grows exponentially with the network size. These drawbacks are overcame by our framework, as it considers the dynamics of all nodes at any level of resolution
and it provides a HOI measure for each level: the OIR of all processes as collective network-wise measure, the local OIR between two processes and the rest of the network as link-wise measure, and the OIR-gradient between one process and the rest of the network as node-wise measure. Ultimately, merging these different types of information through the standard formalism of network depiction leads to provide a compact, yet comprehensive representation of high-order interactions as networks.

The potentiality of the new framework was first demonstrated in a theoretical example of a dynamic network with dynamics modeled by a VAR process, for which exact calculation of the HOI measures was provided under different configurations of the network. Through this example we have shown how higher-order behaviors of a network system, even as simple as a first-order linear process, do not necessarily rely on higher-order mechanisms \cite{rosas2022disentangling}. In fact, simple pairwise linear couplings between random processes can give rise either to fully redundant HOIs related to common driver effects, to fully synergistic HOIs related to common target effects, or to mixed redundant and synergistic HOIs related to coexisting common driver, cascade and common target effects \cite{mackinnon2021unification}. The latter situation, which is very common in complex networks with intricate (though low-order) connections such as those found in computational physiology \cite{bashan2012network, javorka2017basic,mijatovic2024assessing}, underlines the importance to explore higher-order behaviours at different levels of resolution. In particular, we have shown that the coexistence of redundant and synergistic HOIs can be often elicited only by
going beyond the global assessment provided by the OIR through the use of its link-wise or node-wise versions (see, e.g., Fig. \ref{simu_res}c, f). The implementation of the HOI measures in short-length realizations of the processes, complemented by the statistical validation based on surrogate/bootstrap time series, have documented the feasibility of their estimation in practical contexts.

The proposed framework was then tested in a physiological application, demonstrating that HOI networks offer additional and sometimes complementary information compared to standard networks which exclusively depict LOIs focusing on individual node activity and pairwise functional connectivity. Specifically, in the observed cardiovascular networks, we documented the existence of heterogeneous HOIs among the beat-to-beat variability of physiological processes monitored at rest and during stress, identifying respectively the heart period and the diastolic pressure, and the cardiac output and the peripheral resistance, as the processes evoking lower net redundancy during postural stress and higher net synergy during mental stress, also identifying the putative neuro-mechanical mechanisms associated with network reconfiguration \cite{cohen2002short,thomas2011neural,javorka2017causal,mijatovic2024assessing,gambuzza2021stability}. These results have obvious physiological relevance related to how physiological mechanisms can be interpreted in terms of the high-order behaviors that they produce in cardiovascular oscillations, but can also have clinical importance related to the extraction of HOI-based biomarkers reflecting system-wise pathological alterations \cite{ivanov2016focus}, as well as practical relevance thanks to the fact that multivariate physiological time series are nowadays easy to record in daily life settings and can be exploited to assess the cardiovascular control in several contexts including stress monitoring \cite{zanetti2021multilevel}, aging \cite{romero2021network}, or exercise \cite{iellamo2001neural}.

Importantly, thanks to its information-theoretic foundation and to the data-driven implementation, our framework is very general and can be applied virtually to any multivariate dataset featuring time series with temporal auto- and cross-correlations.
In fact, the applicability of our approach for depicting HOI networks is broad and extends far beyond the description of cardiovascular networks, encompassing dynamic networks in fields where the node activity can be  represented by random processes, such as neuroscience, electronics, climatology, social sciences, finance, and others \cite{alvarez2021evolutionary,antonacci2021measuring,luppi2024information, gambuzza2021stability, scagliarini2023gradients}.
Moreover, the framework is highly flexible and lends itself to extensions that will  further broaden its applicability, including its recent and ongoing model-free \cite{ehrlich2024partial}, time-varying \cite{antonacci2023time} and spectral  \cite{sparacino2024measuring} formulations, that will allow to capture complex high-order behaviors and dissect non-stationary and frequency-specific HOIs that are relevant to a big variety of biological and engineered networks.

\section*{Software availability}
The Matlab functions that perform the HOIs framework are available for free download at: \\ https://github.com/mijatovicg/HOIs as networks.

\bibliography{HOInet}

\end{document}